\documentclass[a4paper,11pt,twoside]{article}
\usepackage[english]{babel}
\makeatletter
\usepackage{geometry,graphicx,color}
\usepackage{amsmath,amssymb}
\usepackage{hyperref,cite,url}
\usepackage{enumitem}
\numberwithin{equation}{section}
\newcommand{\institute}[1]{\newcommand{\@institute}{#1}}
\renewcommand{\maketitle}{
\vspace*{0.5\baselineskip}
{
\center\LARGE\noindent\@title\par
}%
\vspace{1.5\baselineskip}
{
\center\normalsize\noindent\ignorespaces\@author\par
}%
\vspace{0.5\baselineskip}
{
\center\normalsize\ignorespaces\@institute\par
}%
\vspace{2\baselineskip}
}%
\let\OLDthebibliography\thebibliography%
\renewcommand\thebibliography[1]{%
\OLDthebibliography{#1}%
\setlength{\parskip}{0pt}%
\setlength{\itemsep}{0pt plus 0.3ex}%
}%
\begin{document}
\title{$\kappa$-Poincar\'e invariant orientable field theories at 1-loop}
\author{Timoth\'e Poulain, Jean-Christophe Wallet}
\institute{%
\textit{Laboratoire de Physique Th\'eorique, B\^at.\ 210\\%
CNRS and Universit\'e Paris-Sud 11,  91405 Orsay Cedex, France}\\%
e-mail: \href{mailto:timothe.poulain@th.u-psud.fr}{\texttt{timothe.poulain@th.u-psud.fr}}, \href{mailto:jean-christophe.wallet@th.u-psud.fr}{\texttt{jean-christophe.wallet@th.u-psud.fr}}\\[1ex]%
}%
\maketitle
\begin{abstract} 
We consider a family of $\kappa$-Poincar\'e invariant scalar field theories on 4-d $\kappa$-Minkowski space with quartic orientable interaction, that is for which $\phi$ and its conjugate $\phi^\dag$ alternate in the quartic interaction, and whose kinetic operator is the square of a $U_\kappa(iso(4))$-equivariant Dirac operator. The formal commutative limit yields the standard complex $\phi^4$ theory. We find that the 2-point function receives UV linearly diverging 1-loop corrections while it stays free of IR singularities that would signal occurrence of UV/IR mixing. We find that all the 1-loop planar and non-planar contributions to the 4-point function are UV finite, stemming from the existence of the particular estimate for the propagator partly combined with its decay properties at large momenta, implying formally vanishing of the beta-functions at 1-loop so that the coupling constants stay scale-invariant at 1-loop.

\end{abstract}
\newpage
\section{Introduction}

One of the most studied non-commutative spaces is the $\kappa$-Minkowski 
space-time \cite{MR1994}, \cite{luk1} which often appears to be a reasonable candidate for a quantum space-time which may show up in a proper description of quantum gravity. Emergence of noncommutative structures are expected to likely occur at the Planck scale \cite{Doplich1} where quantum gravity regime becomes relevant. Another argument supporting this belief is the observation \cite{maatz} that (2+1)-d quantum gravity with matter leads upon integrating out the gravitational degrees of freedom to a Noncommutative Field Theory (NCFT) on a noncommutative (quantum) space which reduces to Minkowski space in the low-energy/commutative limit, the Planck scale playing basically the role of a noncommutativity parameter. The resulting NCFT is invariant under the transformations pertaining to a Hopf algebra which basically plays the role of the deformed symmetries of the quantum space. Albeit this feature is obtained only within a (2+1)-d framework which greatly takes advantage of the topological nature of (2+1)-d gravity, it is plausible to assume that a somehow similar structure also prevails in a (3+1)-d situation. In this respect, studying NCFT on $\kappa$-Minkowski space, with $\kappa$-Poincar\'e Hopf algebra modeling 
the space(-time) quantum symmetries, is of great interest.\bigskip

Recall that the $\kappa$-Minkowski noncommutative space(-time) has been properly described a long time ago in \cite{MR1994} by exhibiting the Hopf algebra bicrossproduct structure of the $\kappa$-Poincar\'e quantum algebra \cite{luk1} which co-acts covariantly on it. It can be viewed, informally, as 
the enveloping algebra of the Lie algebra
\begin{equation}\label{algebra}
[x_0,x_i]=\frac{i}{\kappa} \ x_i, \ \ [x_i,x_j]=0, \ \ i,j=1,\cdots,3
\end{equation}
where the deformation parameter $\kappa$, has mass dimension equal to 1. Algebraic structures of (or related to) $\kappa$-Minkowski space and $\kappa$-Poincar\'e algebra, in particular from the viewpoint of quantum groups \cite{leningrad} as well as twists deformations, have been widely explored resulting in a huge literature on the subject. For a comprehensive recent review, see for instance \cite{luk2} (and references therein). From a phenomenological viewpoint, observable consequences related to the very structures of $\kappa$-Minkowski and $\kappa$-Poincar\'e have been carefully discussed in a lot of works, in particular in connection with Doubly Special Relativity, modified dispersion relations and relative locality \cite{ame-ca1,reloc}.\bigskip

NCFT on $\kappa$-Minkowski space have received a lot of interest for the last decades. See e.g \cite{ital-1}-\cite{hrvat-1} and references therein. It turns out that most of the studies deal with their classical properties while their quantum properties were poorly explored until recently \cite{mercati1}, \cite{PW2018}, apart from the work \cite{gross-whl} where the issue of UV/IR mixing was examined within $\kappa$-Poincar\'e invariant scalar field theories and was found to possibly occur. Unfortunately, the technical complexity of the perturbative computations stemming from the very structure of the star product defining the $\kappa$-deformation, which was built in \cite{star-spunz}, has precluded further extension of these primary investigations. In a recent paper, we used \cite{PW2018} a simple adaptation of the Wigner-Weyl quantization scheme combined with structural properties of the convolution algebra of the affine group $\mathbb{R}\ltimes\mathbb{R}^3$ (formathematical details on related harmonic analysis see e.g \cite{dana}), a non unimodular Lie group, to obtain a natural star product for 4-d $\kappa$-Minkowski space \cite{PW2018}. This star product was also considered in somehow different contexts in \cite{DS,matas}. This star product proved very convenient to explore quantum properties of related NCFT, thanks to its simple expression together with its properties stemming from Lie group harmonic analysis which make him well suited to investigate quantum field theories. Note that a somehow similar construction mainly based on the machinery of harmonic analysis on Lie groups controls entirely the works \cite{vitwal1}-\cite{JCW2016} dealing with $\mathbb{R}^3_\lambda$, a deformation of $\mathbb{R}^3$, and related NCFT, in which however the compact (hence unimodular) Lie group $SU(2)$ replaces the above non unimodular affine group, leading to nice simplifications stemming from the Peter-Weil theorem.\bigskip

In \cite{PW2018}, we used the above natural star product to study general 1-loop properties of a wide family of $\kappa$-Poincar\'e invariant(complex) scalar NCFT on 4-d $\kappa$-Minkowski whose commutative limit is the usual massive $\phi^4$ theory. It turns out that the requirement of $\kappa$-Poincar\'e invariance implies that the (Lebesgue) integral contained in the action functional behaves as a twisted trace with respect to the star product, hence defining a KMS weight \cite{kuster} on the algebra (of fields) modeling the $\kappa$-Minkowski space which may be viewed as the property replacing the usual cyclicity of the trace in the action functional. The corresponding modular group together with Tomita modular operator \cite{takesaki} and related possible physical implications related to a global (observer-independent) time \cite{ConRove} were discussed in \cite{PW2018}. The perturbative investigations were limited to the UV and IR behavior of the 2-point functions, showing in particular the salient role played by the twist automorphism related to the twisted trace in the control of the UV behavior of the contributions to the 2-point functions. It was also found that the UV/IR mixing does not necessarily appear in these NCFT.\bigskip

As a next step following \cite{PW2018}, we consider in this paper a family of $\kappa$-Poincar\'e invariant scalar field theories on 4-d $\kappa$-Minkowski space involving the most general quartic orientable interaction term. Recall that in the liturgy of NCFT, the interaction is called orientable \cite{vign-sym} whenever the fields, says $\phi$, and its conjugate, says $\phi^\dag$, alternate in the quartic interaction terms{\footnote{Orientable NCFT on the Moyal spaces $\mathbb{R}^4_\theta$ and on $\mathbb{R}^3_\lambda$ have been studied in detail paying attention to their quantum and renormalizability properties in \cite{Wallet:2007c}}}. The kinetic operator will be assumed to be the square of the $U_\kappa(iso(4))$-equivariant Dirac operator constructed in \cite{dandrea2006}, which was also examined in \cite{PW2018}. The family of classical NCFT which  satisfies a reality condition with respect to a natural Hilbert product, is indexed by two real dimensionless coupling constants together with a parameter with mass dimension and reduces in the formal commutative limit $\kappa\to\infty$ to the standard complex $\phi^4$ theory. The computation of the 1-loop contributions to the 2- and 4- point functions reduces to control the convergence of time-like integrals which is achieved by using a regulator inspired by the form of the generators of a subalgebra of the Hopf $\kappa$-Poincar\'e algebra. We find that the 2-point function receives UV linearly diverging 1-loop corrections. Besides, it does not involve IR singularities that would signal occurrence of UV/IR mixing. We also find that the 1-loop planar and non-planar contributions to the 4-point function are UV finite, as a consequence of the existence of the particular estimate for the propagator combined with its decay properties at large momenta. Since the coupling constants receive only finite renormalization, they will not depend on a (mass) scale at this 1-loop order, i.e they are scale-invariant which signals that the beta functions are zero at 1-loop.\bigskip

The paper is organized as follows. In the section \ref{section2}, we collect the useful properties underlying the scalar NCFT we consider.  In the Section \ref{section3}, we present the computation of the 1-loop contributions to the 2-point function as well as the regularization used to deal with the various integrals. In the Section \ref{section4}, the analysis of the 1-loop contributions to the 4-point function is carried out. In the Section \ref{section5}, we discuss the results and we conclude.\\\

\paragraph{Notations:} In the following, space-like (resp. time-like) coordinates are denoted as usual by Latin indices $i,j,\cdots=1,2,3$ (resp. $0$ indices);  Greek indices ($\mu,\nu,\cdots$) run from $0$ to $3$. For any 4-vector $x$, we set $x=(x_\mu)=(x_0,\vec{x})$ and $x.y=x_\mu y^\mu=x_0y_0+\vec{x}\vec{y}$, thus working with Euclidean signature. Einstein summation convention for repeated indices is assumed.\\
We define the Fourier transform of $f\in L^1(\mathbb{R}^4)$ by $(\mathcal{F}f)(p):=\int d^4x\ e^{-i(p_0x_0+\vec{p}.\vec{x})}f(x)$ with inverse $\mathcal{F}^{-1}$ and denote by $\bar{f}$ its complex conjugate. $\mathcal{S}_c$ is the space of Schwartz functions on $\mathbb{R}^4=\mathbb{R}\times\mathbb{R}^3$ with compact support in the first variable. 
\section{\texorpdfstring{$\kappa$}{k}-Poincar\'e invariant orientable scalar field theories}\label{section2}
In this section, we collect the main properties of the family of $\kappa$-Poincar\'e invariant scalar field theories which we consider. The corresponding action functional is given by
\begin{equation}
S_\kappa(\phi^\dag,\phi)=S^\text{kin}_\kappa(\phi^\dag,\phi)+S^\text{int}_\kappa(\phi^\dag,\phi),\label{kaction}
\end{equation}
where the kinetic part is defined by
\begin{align}\label{kinetic-term}
S^\text{kin}_\kappa(\phi^\dag,\phi)&=\frac{1}{4}\langle \phi, (K_\kappa+m^2)\phi \rangle+\frac{1}{4}\langle \phi^\dag,(K_\kappa+m^2)\phi^\dag \rangle
\end{align}
in which the kinetic operator ${K}_\kappa$ is a self-adjoint differential operator{\footnote{for any $\phi$ in the domain of ${K}_\kappa$, dense in the Hilbert space $\mathcal{H}\simeq L^2(\mathbb{R}^4)$ defined in \cite{PW2018}.}} satisfying
\begin{equation}\label{pseudoper}
(K_\kappa \phi)(x)=\int \frac{d^4p}{(2\pi)^4} d^4y\ {K}_\kappa(p)\phi(y)e^{ip\cdot(x-y)},
\end{equation}
and ${K}_\kappa(p)$ is a (real) function of the momenta to be characterized in a while, together with the Hilbert product $\langle.,.\rangle$ used in \eqref{kinetic-term}.\\
The interaction term is assumed to be 
\begin{equation}\label{interaction}
S^\text{int}_\kappa(\phi^\dag,\phi)=g_1\int d^4x\ (\phi^\dag\star\phi\star\phi^\dag\star\phi)(x)+g_2\int d^4x\ (\phi\star\phi^\dag\star\phi\star\phi^\dag)(x),
\end{equation}
where $g_1,g_2\in\mathbb{R}$. Here, $\phi$ is a complex-valued field, $^\dag$ denotes the natural involution which comes along with the star product stemming from the Weyl-type quantization \cite{PW2018}. Hence, $\phi^\dag$ is not the complex conjugate of $\phi$, denoted by $\bar{\phi}$.\\

Let us first recall some basic properties of the star product underlying the present study.\\As explained in \cite{DS,matas,PW2018}, a convenient star product describing the $\kappa$-Minkowski space together with the related involution can be defined by
\begin{equation}
f\star g:=\mathcal{F}^{-1}(\mathcal{F}f\circ\mathcal{F}g),\ \ f^\dag:=\mathcal{F}^{-1}((\mathcal{F}f)^*)\label{basic-def1},
\end{equation}
for any $f,g\in\mathcal{F}(\mathcal{S}_c)$, where $\circ$ (resp. $^*$) denotes the right-convolution product (resp. the canonical involution) equipping the convolution algebra of the affine group{\footnote{\label{note-convol}The convolution algebra $L^1(\mathcal{G})$ is the set of integrable $\mathbb{C}$-valued functions on the (non unimodular) group $\mathcal{G}:=\mathbb{R}^+_{/0}\ltimes_\phi\mathbb{R}^3$, with $\phi:\mathbb{R}^+_{/0}\to\textrm{Aut}(\mathbb{R}^3)$ the adjoint action of $\mathbb{R}^+_{/0}$ on $\mathbb{R}^3$. The convolution product w.r.t to the right-invariant Haar measure is defined, for any $t\in\mathcal{G}$ and $f, g\in{L^1(\mathcal{G})}$, by $(f\circ g)(t):=\int_{\mathcal{G}}d\nu(s)\ f(ts^{-1})g(s)$ with involution $f^*(t):=\bar{f}(t^{-1})\Delta_{\mathcal{G}}(t)$ and  modular function $\Delta_{\mathcal{G}}$, turning $L^1(\mathcal{G})$ into a $^*$-algebra. See e.g \cite{dana}.}}, $\mathbb{R}\ltimes\mathbb{R}^3$. This gives rise to
\begin{align}
(f\star g)(x)&=\int \frac{dp^0}{2\pi}dy_0\ e^{-iy_0p^0}f(x_0+y_0,\vec{x})g(x_0,e^{-p^0/\kappa}\vec{x})  \label{starpro-4d},\\
f^\dag(x)&= \int \frac{dp^0}{2\pi}dy_0\ e^{-iy_0p^0}{\bar{f}}(x_0+y_0,e^{-p^0/\kappa}\vec{x})\label{invol-4d},
\end{align}
with $f\star g\in\mathcal{F}(\mathcal{S}_c)$ and $f^\dag\in\mathcal{F}(\mathcal{S}_c)$.\\Eqns. \eqref{starpro-4d} and \eqref{invol-4d} can be extended \cite{DS} to a multiplier algebra involving the smooth functions on $\mathbb{R}^4$ with standard polynomial bounds together with all their derivatives, including $x_0$, $\vec{x}$ and the unit function. From \eqref{starpro-4d} and \eqref{invol-4d}, a simple calculation yields
\begin{equation}\label{def-relations}
x_0\star x_i=x_0x_i+\frac{i}{\kappa}x_i,\ \ x_i\star x_0=x_0x_i,\ \ x_\mu^\dag=x_\mu,
\end{equation}
$i=1,\cdots,3$, $\mu=0,\cdots,3$, which is consistent with the defining relation \eqref{algebra} for the $\kappa$-Minkowski space. From now on, we denote the algebra describing the $\kappa$-Minkowski space by $\mathcal{M}_\kappa$.\\

A natural integration measure for $\kappa$-Minkowski is provided by the right-invariant Haar measure $d\nu$ related to the convolution algebra of the affine group (see footnote \ref{note-convol}). It reduces to the usual Lebesgue measure when expressed with the momentum variables appearing in the parametrization of the affine group elements \cite{DS,matas,PW2018} so that $\mathcal{F}:d\nu(p_0,\vec{p})=d^4p\to d^4x$.\\

A convenient Hilbert product that will be used \cite{PW2018} to define reality condition for the action functionals is given by
\begin{equation}
\langle f,g\rangle:=\int d^4x\ \left(f^\dag\star g\right)(x),\label{hilbert-product}
\end{equation}
for any $f,g\in\mathcal{M}_\kappa$. Reality condition for the kinetic term of the form $\langle f,K_\kappa f\rangle$ is automatically verified since the kinetic operator $K_\kappa$ that we will consider is self-adjoint. Indeed, in this case one has $\langle f,K_\kappa f\rangle=\langle K_\kappa f,f\rangle$. Besides, the reality of the interaction term  \eqref{interaction} is apparent by simply noticing that \eqref{interaction} can be expressed in term of the Hilbert product as $S^\text{int}_\kappa(\phi^\dag,\phi)=g_1\langle\phi^\dag\star\phi,\phi^\dag\star\phi\rangle\ +\ g_2\langle\phi\star\phi^\dag ,\phi\star\phi^\dag \rangle$.\\

Besides, it is well known that any action functional of the form
\begin{equation}
S_\kappa(\phi)=\int d^4x\ \mathcal{L}(\phi)
\end{equation}
as the one given by \eqref{kaction} is invariant under the action of the $\kappa$-Poincar\'e transformations. Indeed, recall that one can easily verify that one has
\begin{equation}\label{invarquant}
h\blacktriangleright S_\kappa(\phi):=\int d^4x\ h\triangleright\mathcal{L}(\phi)=\epsilon(h)S_\kappa(\phi),
\end{equation}
for any $h$ in the $\kappa$-Poincar\'e Hopf algebra $\mathcal{P}_\kappa$, where $\epsilon: \mathcal{P}_\kappa\to\mathbb{C}$ is the co-unit of $\mathcal{P}_\kappa$. This latter is defined by $\epsilon(P_0)=\epsilon(P_i)=\epsilon(M_i)=\epsilon(N_i)=0,\  \epsilon(e^{-P_0/\kappa})=1 $, where $P_\mu$, $M_i$ and $N_i$ denote respectively the momenta, the rotations and the boosts.\\
Useful formulas which will be needed in the ensuing analysis are given by
\begin{equation}\label{int-form1}
\int d^4x\ (f\star g^\dag)(x)=\int d^4x\ f(x){\bar{g}}(x),\ \ \int d^4x\ f^\dag(x)=\int d^4x\ {\bar{f}}(x),
\end{equation}
while, defining
\begin{equation}\label{twistoperator}
\left(\sigma\triangleright f\right)(x_0,\vec{x}):=\left(e^{i\frac{3\partial_0}{\kappa}}\triangleright f\right)(x_0,\vec{x})=f(x_0+\frac{3i}{\kappa}, \vec{x}),
\end{equation}
one can verify that
\begin{equation}\label{twistrace}
\int d^4x\ (f\star g)(x)=\int d^4x\ \left((\sigma\triangleright g)\star f\right)(x).
\end{equation}
This last relation signals that the Lebesgue integral defines a twisted trace \cite{DS,matas,PW2018}, hence implying a loss of cyclicity with respect to the star product. The corresponding twist is given by $\sigma$, eqn. \eqref{twistoperator}. Recall that it defines an automorphism of the noncommutative algebra modeling the $\kappa$-Minkowski space. One can easily check that 
\begin{equation}
\sigma(f\star g)=\sigma(f)\star\sigma(g),
\end{equation}
together with
\begin{equation}
\sigma(f^\dag)=(\sigma^{-1}(f))^\dag,
\end{equation}
the latter formula signaling that $\sigma$ is not a $^*$-automorphism. This type of automorphim is known in the mathematical literature as a regular automorphim. It occurs in the framework of twisted spectral triples \cite{CM2008}. It turns out that any (positive) functional built from the twisted trace \eqref{twistrace}, i.e $\omega(f)=\int d^4x\ f(x)$ with $\omega(f)\ge0$ for $f\ge0$, defines a KMS weight. Some corresponding consequences on a NCFT described by an action of the form $S(\phi)=\int d^4x\ \mathcal{L}(\phi)$ together with relevant mathematical details can be found in \cite{PW2018}.\\

We assume that the kinetic operators $K_\kappa$ considered in the ensuing analysis is the square of the $U_\kappa(iso(4))$-equivariant Dirac operators constructed a long ago in \cite{dandrea2006}. Some properties of the quantum behavior of scalar NCFT it is involved in has been examined in \cite{PW2018}. \\
The equivariant Dirac operator is defined by
\begin{equation}\label{equivdirac}
D^{eq}_0=\frac{\kappa}{2}\mathcal{E}^{-1}(1-\mathcal{E}^2)-\frac{1}{2\kappa}\mathcal{E}^{-1}\vec{P}^2 \ \ ,\ D^{eq}_i=\mathcal{E}^{-1}P_i ,\ \ i=1,\cdots,3.
\end{equation}
Accordingly, the kinetic operator is given by
\begin{equation}\label{equi-dirac}
K^{eq}_\kappa(P_\mu):= \left(D^{eq}_0\right)^2+\sum_{i=1}^3 D^{eq}_iD^{eq}_i=\mathcal{C}_\kappa(P_\mu)+\frac{1}{4\kappa^2}\mathcal{C}_\kappa(P_\mu)^2,
\end{equation}
where $\mathcal{C}_\kappa(P_\mu)$ is the first Casimir of the $\kappa$-Poincar\'e algebra. This latter is given, in the Majid-Ruegg basis \cite{MR1994}, by
\begin{equation}\label{Casimir}
\mathcal{C}_\kappa(P_\mu)=4\kappa^2\sinh^2\left(\frac{P_0}{2\kappa}\right) + e^\frac{P_0}{\kappa} P_iP^i,
\end{equation}
where $P_\mu\in\mathcal{P}_\kappa$ acts on $\mathcal{M}_\kappa$ by $P_\mu=-i\partial_\mu$.\\
Note that \eqref{equivdirac} define self-adjoint operators since one has
\begin{equation}
\langle \mathcal{D}_\mu f,g \rangle=\langle f,\mathcal{D}_\mu g \rangle\label{dequiv-selfadj}
\end{equation}
for any $f,g\in\mathcal{M}_\kappa$. This can be easily shown by using the known duality of $^*$-algebras between $\mathcal{M}_\kappa$ and $\mathcal{T}_\kappa$, the deformed translation Hopf subalgebra of $\mathcal{P}_\kappa$ generated by $P_\mu$ and $\mathcal{E}$ which entails 
\begin{equation}
(t\triangleright f)^\dag=S(t)^\dag\triangleright f^\dag
\end{equation}
for any $t\in \mathcal{T}_\kappa$ and any $f\in\mathcal{M}_\kappa$, together with $S(P_0)=-P_0,\ S(\mathcal{E})=\mathcal{E}^{-1},\  S(P_i)=-\mathcal{E}^{-1}P_i$, where 
$S:\mathcal{P}_\kappa\to\mathcal{P}_\kappa$ is the antipode of $\mathcal{P}_\kappa$ and
\begin{align}
P_0\triangleright(f\star g)&=(P_0\triangleright f)\star g+f\star(P_0\triangleright  g )\label{deriv-twist2},\\
P_i\triangleright(f\star g)&=(P_i\triangleright f)\star g+(\mathcal{E}\triangleright f)\star (P_i\triangleright g)\label{deriv-twist1}.
\end{align}
Note that one also has
\begin{equation}\label{parpart}
\langle \mathcal{D}_\mu\phi, \mathcal{D}^\mu\phi \rangle=\langle \phi, \mathcal{D}_\mu\mathcal{D}^\mu\phi \rangle=\langle \phi,K_\kappa(P_\mu)\phi\rangle,
\end{equation}
which may be used to simplify the expression of the action functional. From \eqref{dequiv-selfadj}, \eqref{parpart}, one concludes that $K^{eq}_\kappa(P_\mu)$ is self-adjoint, as annouced at the beginning of this section.\\

The corresponding family of kinetic operators, when supplemented by a mass term, $m\leq\kappa$, can be factorised as
\begin{equation}\label{equi-kin}
K_\kappa(P_\mu)+m^2 = \frac{e^{2P_0/\kappa}}{4\kappa^2}\left(\vec{P}^{\hspace{2pt}2}+\kappa^2\mu^2_{+}\right)\left(\vec{P}^{\hspace{2pt}2}+\kappa^2\mu^2_{-}\right),
\end{equation}
where the (positive) functions $\mu^2_{+}$ and $\mu^2_{-}$ are given by
\begin{equation}\label{eq-miou}
\mu^2_{\pm}(m,P_0):=1\pm 2 e^{-P_0/\kappa}\sqrt{1-\left(\frac{m}{\kappa}\right)^2}+e^{-2P_0/\kappa},
\end{equation}
from which \eqref{equi-kin} can be inverted.\\

Finally, by using \eqref{pseudoper}, \eqref{hilbert-product} and \eqref{twistrace}, the kinetic term \eqref{kinetic-term} can be cast into the form
\begin{align}\label{kinetic-term1}
S^\text{kin}_\kappa({\phi}^\dag,\phi)&=\frac{1}{4}\langle \phi, (K_\kappa+m^2)\phi \rangle+\frac{1}{4}\langle \phi^\dag,(K_\kappa+m^2)\phi^\dag \rangle\nonumber
\\&=\frac{1}{4}\int d^4x\left(\phi^\dag\star(1+\sigma^{-1})(K_\kappa+m^2)\phi \right)(x),
\end{align}
while, by combining the integral representation \eqref{starpro-4d} of the star product into \eqref{kaction}, one easily shows that any $\kappa$-Poincar\'e invariant NCFT can be represented, as expected, as a standard albeit non-local quantum field theory. In momentum space, the corresponding action takes the form $S_\kappa(\bar{\phi},\phi)=S^\text{kin}_\kappa(\bar{\phi},\phi)+S^\text{int}_\kappa(\bar{\phi},\phi)$, where $\phi$ and $\bar{\phi}$ now denote the Fourier transform of the classical fields, with
\begin{align}
S^\text{kin}_\kappa(\bar{\phi},\phi)=&\ \frac{1}{2}\int \frac{d^4p}{(2\pi)^4}\ \bar{\phi}(p)\phi(p)\mathcal{K}(p),\label{decadix-77}\\
\mathcal{K}(p):=&\ \frac{1}{2}\left(1+e^{-3p^0/\kappa}\right)\left(K_\kappa(p)+m^2\right),\label{decadix-7}
\end{align}
where $K_\kappa(p)+m^2$ can be read off from \eqref{equi-kin} and
\begin{align}
S^\text{int}_\kappa(\bar{\phi},\phi)=&\ \int \left[\prod_{i=1}^4\frac{d^4p_i}{(2\pi)^4}\right] \bar{\phi}(p_1)\phi(p_2)\bar{\phi}(p_3)
\phi(p_4) V(p_1,p_2;p_3,p_4) ,\\
V(p_1,p_2;p_3,p_4):=&\ (g_1+g_2e^{3p_1^0/\kappa}) 
(2\pi)^4\delta\left(p_2^0-p_1^0+p_4^0-p_3^0\right)\\
&\ \times\delta^{(3)}\left(\left(\vec{p}_2-\vec{p}_1
\hspace{2pt}\right)e^{p_1^0/\kappa}+\left(\vec{p}_4-\vec{p}_3
\hspace{2pt}\right)e^{p_4^0/\kappa}\right).\label{vertex}
\end{align}
The expression for the vertex function \eqref{vertex} exhibits the usual conservation law for the energy, while the conservation law for the space-like momenta becomes non-linear expressing the non-locality of the quartic interaction as expected from the very structure of the affine group.
In \eqref{decadix-77}-\eqref{vertex}, the mass dimension for the fields and parameters have their canonical mass dimensions, namely $[\phi]=[\phi^\dag]=1$, $[g]=0$ and $[m]=[\kappa]=1$. \\

We now pass to the computation of the 2-point and 4-point functions at the one-loop order.\\

\section{Two-point functions}\label{section3}

One-loop corrections to the 2-point functions is obtained as usual by expanding the generating functional of the connected correlation functions
\begin{equation}
W[\bar{J},J]:=\ln\left(Z[\bar{J},J]\right)=-\ln\left(Z[0,0]\right)+\ln\left(W_0[\bar{J},J]\right)+\ln\left(e^{-W_0}e^{-S_\kappa^\text{int}[\frac{\delta}{\delta J},\frac{\delta}{\delta\bar{J}}]}e^{W_0}\right),\label{correlation}
\end{equation}
in which $Z[\bar{J},J]$ is the partition function defined by
\begin{equation}
\mathcal{Z}_\kappa[\bar{J},J]:=\int d\bar{\phi}d\phi \ e^{-S^\text{kin}_\kappa(\bar{\phi},\phi)-S^\text{int}_{\kappa}(\bar{\phi},\phi)+\int d^4x \ \bar{J}(x)\phi(x) + \int d^4x \ J(x)\bar{\phi}(x)}.\label{partition}
\end{equation}
where $J$ and $\bar{J}$ are source and
\begin{equation}
W_0[\bar{J},J]:=\int \frac{d^4p}{(2\pi)^4} \ \bar{J}(p)P_\kappa(p)J(p),
\end{equation}
is the generating functional of connected correlation functions for the free field theory in which $P_\kappa(p)$ is the propagator, i.e
\begin{equation}
P_\kappa(p)=\mathcal{K}^{-1}(p) 
\end{equation}
and $\mathcal{K}(p)$ is given by \eqref{decadix-7}. Then, the quadratic part of the effective action denoted hereafter by $\Gamma(\phi,\bar{\phi})$ is obtained by retaining the terms in the expansion of \eqref{correlation} up to the first order in the coupling constants and performing an inverse
Legendre transform. One has
\begin{equation}
\Gamma(\phi,\bar{\phi})=\int\ \frac{d^4k}{(2\pi^4)}(\bar{J}(k)\phi(k)+J(k)\bar{\phi}(k))-W(\bar{J},J)\label{legendre},
\end{equation}
with
\begin{equation}
\bar{\phi}(k)=\frac{\delta W[\bar{J},J]}{\delta J(k)},\ \ {\phi}(k)=\frac{\delta W[\bar{J},J]}{\delta \bar{J}(k)}\label{phi-J}.
\end{equation}
The inversion yields at the lowest order in the coupling constants
\begin{equation}
\bar{J}(k)=-\mathcal{K}(k)\bar{\phi}(k)+\mathcal{O}(g_1,g_2),\ \ {J}(k)=-\mathcal{K}(k){\phi}(k)+\mathcal{O}(g_1,g_2)\label{J-phi}.
\end{equation}

This latter, combined with the suitable expansion of \eqref{correlation} together with \eqref{J-phi} and \eqref{legendre} gives rise to the one-loop quadratic part of $\Gamma$ given by
\begin{align}
\Gamma^{(2)}[\bar{\phi},\phi]&=\int\frac{d^4p_3}{(2\pi)^4}\frac{d^4p_4}{(2\pi)^4}\  \bar{\phi}(p_3)\phi(p_4) \Gamma^{(2)}(p_3,p_4),\\
\Gamma^{(2)}(p_3,p_4)&=\int \frac{d^4p_1}{(2\pi)^4}\frac{d^4p_2}{(2\pi)^4}\ P_\kappa(p_1) \delta^{(4)}(p_2-p_1)\left[V_{12;34}+V_{34;12}+V_{32;14}+V_{14;32}\right]\label{gen-2pts}
\end{align}
where we have set 
\begin{equation}
V_{12;34}:=V(p_1,p_2;p_3,p_4).
\end{equation}

Inserting the expression \eqref{vertex} for the vertex function into \eqref{gen-2pts}, one finds after standard calculation that the one-loop quadratic part of $\Gamma$ can be cast into the form
\begin{equation}
\Gamma^{(2)}[\bar{\phi},\phi]=\int\ \frac{d^4k}{(2\pi)^{4}}\bar{\phi}(k)(\omega_1+\omega_2e^{-3k_0/\kappa})\phi(k)\label{gamma2-gene},
\end{equation}
with
\begin{equation}
\omega_1=\int\frac{d^4k}{(2\pi)^{4}}\ (3g_2+g_1e^{-3k_0/\kappa})P_\kappa(k),\ \ \omega_2=\int\frac{d^4k}{(2\pi)^{4}}\ (3g_1+g_2e^{3k_0/\kappa})P_\kappa(k)\label{omega-start}.
\end{equation}
The expression for the propagator stemming from \eqref{equi-kin}, \eqref{decadix-7} is given by
\begin{equation}
P_{\kappa}(q)=8\kappa^2\frac{e^{-2q^0/\kappa}}{1+e^{-3q^0/\kappa}} \  \frac{1}{\left(\vec{q}^{\hspace{2pt}2}+\kappa^2\mu^2_{+}\right)\left(\vec{q}^{\hspace{2pt}2}+
\kappa^2\mu^2_{-}\right)},\label{propag2}
\end{equation}
where $\mu^2_\pm$ can be read off from \eqref{eq-miou}, namely
\begin{equation}
\mu_{\pm}^2(p_0)=1\pm2e^{-p_0/\kappa}\sqrt{1-\frac{m^2}{\kappa^2}}
+e^{-2p_0/\kappa}\label{miou-final}.
\end{equation}
Now, by further combining \eqref{omega-start} with \eqref{propag2} and \eqref{miou-final}, one easily realizes that the spatial integrals in \eqref{omega-start} are convergent and can be exactly performed. Namely, upon using successively the two relations
\begin{align}
\frac{1}{A^aB^b}=\frac{\Gamma(a+b)}{\Gamma(a)\Gamma(b)} & \int_0^1 du \ \frac{u^{a-1}(1-u)^{b-1}}{\left(uA+(1-u)B\right)^{a+b}}, \ a,b>0,\label{formule1} \\
\int \frac{d^np}{(2\pi)^n} \frac{1}{(p^2+M^2)^{m}}&=M^{n-2m}\frac{\Gamma(m-n/2)}{(4\pi)^{n/2} \Gamma(m)}, \ m>n/2> 0,\label{formule2}
\end{align}
where $\Gamma(z)$ is the Euler gamma function, one finds after standard calculation and defining as in \cite{PW2018}
\begin{equation}
y=e^{-k_0/\kappa}\label{rapidity},
\end{equation}
that $\omega_1$ and $\omega_2$ can be cast into the form
\begin{equation}
\omega_j=\frac{\kappa^3}{4\pi^2\sqrt{\kappa^2-m^2}}\int dy\  \Phi_j(y)(\sqrt{\mu_+^2(y)}-\sqrt{\mu_-^2(y)}),\ \ j=1,2\label{decadix-develour}
\end{equation}
in which
\begin{equation}
\Phi_1(y)=\frac{3g_2-g_1}{1+y^3}+g_1,\ \ \Phi_2(y)=\frac{3g_1-g_2}{1+y^3}+\frac{g_2}{y^3}\label{grandphi},
\end{equation}
with $\mu_\pm(y)$ still given by \eqref{miou-final} and it is implicitly assumed that the integrals over $y$, which are divergent, are regularized as done in \cite{PW2018}, a scheme which we now recall.\\

We first note that the integration variable $y$, which by the way simplifies the computations, is linked to (some power) of the twist operator \eqref{twistoperator}.  Recalling that the deformed translation Hopf subalgebra of $\mathcal{P}_\kappa$ is generated by $P_\mu$ and $\mathcal{E}$, it is natural to interpret $y$ as related to the ``physical" quantity replacing the time-like momenta, says $q^0$, in the NCFT. This is in some sense apparent by considering the term 
\begin{equation}\label{gen-energy}
K_0(\kappa):=\kappa(1-e^{-q^0/\kappa})=\kappa(1-y),\ \ K_0(\kappa)\xrightarrow[\kappa\to\infty]{}q^0,
\end{equation}
appearing in the Casimir operator \eqref{Casimir}. Observe that this latter can be written as $\mathcal{C}_\kappa(q_0)=e^{q_0/\kappa}(\kappa^2(1-e^{-q_0/\kappa})^2+\vec{q}^{\ 2})$ so that $\kappa^2(1-e^{-q_0/\kappa})^2$ may be viewed as playing the role of $q_0^2$ in the $\kappa$-deformed version of the Casimir operator. According to this remark, it is more natural to impose any cut-off directly on $K_0(\kappa)$, says 
\begin{equation}
\vert K_0(\kappa)\vert\leq\Lambda_0.\label{decadix}
\end{equation}
However, because of the relation \eqref{gen-energy}, the constraint \eqref{decadix} necessarily implies the appearence of a cut-off for $q^0$ and $y$ as well. Assuming $\vert q^0\vert\leq M(\Lambda_0)$, one finds
\begin{equation}
\kappa(1-e^{\frac{M}{\kappa}})\le K_0\le\kappa(1-e^{-\frac{M}{\kappa}}),
\end{equation}
leading to the identification
\begin{equation}
M(\Lambda_0)=\kappa\ln(1+\frac{\Lambda_0}{\kappa}),\ \ M(\Lambda_0)\xrightarrow[\kappa\to\infty]{}\Lambda_0.
\end{equation}
It follows that
\begin{equation}\label{cutygrec}
(1+\frac{\Lambda_0}{\kappa})^{-1}\le y\le1+\frac{\Lambda_0}{\kappa}.
\end{equation}
Hence, the $y$-integrals are understood to be regularized as
\begin{equation}\label{reguly}
\int_0^\infty dy := \lim_{\Lambda_0\to\infty}\int_{(1+\frac{\Lambda_0}{\kappa})^{-1}}^{(1+\frac{\Lambda_0}{\kappa})} dy.
\end{equation}
We will use the latter condition to regularize the $y$-integrals appearing in the computation of the 2- and 4-point functions.\\

Applying the above regularization to \eqref{decadix-develour} yields
\begin{equation}
\omega_j=\frac{\kappa^3}{4\pi^2\sqrt{\kappa^2-m^2}}\int_{(1+\frac{\Lambda_0}{\kappa})^{-1}}^{(1+\frac{\Lambda_0}{\kappa})} dy\  \Phi_j(y)(\sqrt{\mu_+^2(y)}-\sqrt{\mu_-^2(y)}),\ \ j=1,2\label{decadix-regular}
\end{equation}
where $\Phi_j(y)$ are still given by \eqref{grandphi}. By making use of elementary tools of real analysis, one easily infers that
\begin{equation}
\omega_j(\Lambda_0)=\frac{g_j\kappa}{2\pi^2}\Lambda_0+F_j(\kappa),\ \ j=1,2\label{decadix-final}
\end{equation}
where $F_j(\kappa)$, $j=1,2$ are finite contributions for finite $\kappa$ which are given by
\begin{eqnarray}
F_j(\kappa)&=&\frac{g_j\kappa^2}{(2\pi)^2}+\frac{m^2g_j\kappa}{8\pi^2\sqrt{\kappa^2-m^2}}\ln(\frac{\kappa+\sqrt{\kappa^2-m^2}}{\kappa-\sqrt{\kappa^2-m^2}})
+C_j(\kappa)\label{F1},
\end{eqnarray}
for $j=1,2$ with
\begin{eqnarray}
C_1(\kappa)&=&\frac{(3g_2-g_1)\kappa^3}{4\pi^2\sqrt{\kappa^2-m^2}}\int_0^\infty dy\ \frac{\mu_+(y)-\mu_-(y)}{1+y^3},\label{C1}\\
C_2(\kappa)&=&\frac{(3g_1-g_2)\kappa^3}{4\pi^2\sqrt{\kappa^2-m^2}}\int_0^\infty dy\ \frac{\mu_+(y)-\mu_-(y)}{1+y^3}\label{C2}. 
\end{eqnarray}
At this point, some comments are in order.\\
\begin{itemize}
\item We first notice that no IR singularity appears in the 2-point functions which would have signaled the occurrence of UV/IR mixing in the NCFT. Recall that a similar behavior occurs for orientable scalar NCFT defined on Moyal spaces \cite{vign-sym}.
\item From \eqref{decadix-final}, one observes that the 2-point functions have a linear UV divergence while in the commutative case the 2-point function for a scalar $\phi^4$ theory in 4-d is known to have a quadratic UV divergence. This milder UV behavior stems in part from the UV decay properties of the propagator \eqref{propag2}.
\item When $g_1\ne g_2$, \eqref{decadix-final} implies that each of the quadratic mass terms receive different one-loop corrections, namely 
\begin{equation}
\int d^4p\ \bar{\phi}(p)\phi(p)m^2(1+e^{-3p_0/\kappa})\to\int d^4p\ \bar{\phi}(p)\phi(p)(\omega_1+\omega_2e^{-3p_0/\kappa}) 
\end{equation}
with $\omega_1\ne\omega_2$ which can be expected in the absence of symmetry of the action.\\
However, when $g_1=g_2$, it can be easily realized from \eqref{decadix-final} that $\omega_1=\omega_2$ so that the quadratic mass term becomes now stable against (1-loop) radiative corrections. This reflects the fact that the classical action functional \eqref{kaction}-\eqref{interaction} becomes invariant under the symmetry 
\begin{equation}
\phi\to\phi^\dag,\ \ \phi^\dag\to\phi.\label{symmet-nc}
\end{equation}
Eqn. \eqref{symmet-nc} can be expressed in terms of the field variables $\phi,\ \bar{\phi}$ which are better adapted to the perturbative investigations of the action functional \eqref{kaction}-\eqref{interaction}. One easily obtains
\begin{equation}
\phi\to A_\kappa^{-1}\bar{\phi},\ \ \bar{\phi}\to A_\kappa{\phi}\label{symmet-planck},
\end{equation}
where $A_\kappa$ is the unitary intertwiner map $\mathcal{F}(\mathcal{S}_c)\to L^2(\mathbb{R}^4)$ introduced in \cite{PW2018} with
\begin{equation}
(A_\kappa^{-1}\phi)(x)=\int \frac{dp_0}{2\pi}dz_0\ e^{-ip_0z_0}\phi(x_0+z_0,e^{-p_0/\kappa}\vec{x}).
\end{equation}
\end{itemize}

\section{Four-point functions}\label{section4}

\subsection{General structure of the 4-point functions}\label{gen-4pts}
One-loop correction to the 4-point functions are obtained by expanding the generating functional of the connected correlation function \eqref{correlation} up to the second order in the coupling constant.  From a standard computation, one finds that the quartic part of the effective action $\Gamma$ can be written as
\begin{equation}
\Gamma^{(4)}[\bar{\phi},\phi]=\int \left[\prod_{i=3}^6\frac{d^4p_i}{(2\pi)^4}\right]\ \bar{\phi}(p_3)\phi(p_4)\bar{\phi}(p_5)\phi(p_6)\Gamma^{(4)}(p_3,p_4,p_5,p_6),
\end{equation}
where $\Gamma^{(4)}(p_3,p_4,p_5,p_6)$ is given by 

\begin{eqnarray}
\Gamma^{(4)}(p_3,p_4,p_5,p_6)&=&\frac{1}{(2\pi)^8}\int \frac{d^4p_1}{(2\pi)^4}\frac{d^4p_2}{(2\pi)^4}\ P_\kappa(p_1)P_\kappa(p_2)\nonumber\\
&\times & [2V_{16;24}V_{52;31}+2V_{16;24}V_{51;32}+
2V_{14;32}V_{56;21}\nonumber\\
&+&2V_{32;14}V_{56;21}+2V_{12;34}V_{26;51}+2V_{12;34}V_{51;26}\nonumber\\
&+&2V_{14;32}V_{51;26}+V_{14;32}V_{26;51}+V_{32;14}V_{51;26}\nonumber\\
&+&2V_{12;34}V_{56;21}+V_{12;34}V_{21;56}+V_{34;12}V_{56;21}.\label{gamma4}
\end{eqnarray}

By further making use of the following symmetries for the vertex function \eqref{vertex},
\begin{equation}\label{symmetries}
\hat{V}_{12;34}\equiv \hat{V}_{43;21}, \ \ \hat{V}_{12;34}\equiv e^{3(p_3^0-p_4^0)/\kappa}\hat{V}_{34;12},
\end{equation}
together with the ``fusion rules"
\begin{equation}\label{fusion}
\hat{V}_{34;{\bf{21}}}\hat{V}_{{\bf{12}};56}\equiv \hat{V}_{{\bf{12}};56}V_{34;56}, \ \ \hat{V}_{3{\bf{1}};{\bf{2}}6}\hat{V}_{{\bf{1}}4;5{\bf{2}}}\equiv \hat{V}_{{\bf{1}}4;5{\bf{2}}}\hat{V}_{34;56},
\end{equation}
where we defined
\begin{equation}
{V}_{12;34}:=(g_1+g_2e^{3p^0_1/\kappa})\hat{V}_{12;34}
\end{equation}
and
\begin{equation}
\hat{V}_{12;34}:=(2\pi)^4\delta\left(p_2^0-p_1^0+p_4^0-p_3^0\right)
\delta^{(3)}\left(\left(\vec{p}_2-\vec{p}_1
\hspace{2pt}\right)e^{p_1^0/\kappa}+\left(\vec{p}_4-\vec{p}_3
\hspace{2pt}\right)e^{p_4^0/\kappa}\right)\label{v-hat}
\end{equation}
and the bold indices $\bf{1},\ \bf{2}$ corresponds to those momenta which are summed over (with $P_\kappa(p_1)P_\kappa(p_2)$) as in \eqref{gamma4} above, which both hold true in the sense of distributions, one can classify the terms appearing in \eqref{gamma4} into four types of contributions, each family being characterized by the relative position of the contracted momenta appearing in the vertex functions in \eqref{gamma4}. \\
Two among these 4 types can be regarded as planar contributions, for which the fusion rules \eqref{fusion} can be used, unlike the two remaining types which can be interpreted as non-planar contributions. These latter contributions, which show up in NCFT, result from the non linearity of the delta conservation law for the 3-momenta. We set formally
\begin{equation}
\Gamma^{(4)}=\Gamma_1^P+\Gamma_2^P+\Gamma_1^{NP}+
\Gamma_2^{NP}\label{decomp}.
\end{equation}

The two types of planar contributions, denoted above by $\Gamma_1^P$ and $\Gamma_2^P$, can be conveniently written as
\begin{align}
\Gamma^P_1(p_3,p_4,p_5,p_6):=&\ \frac{1}{(2\pi)^8}\hat{V}_{34;56}\times\int \frac{d^4p_1}{(2\pi)^4}\frac{d^4p_2}{(2\pi)^4}\ \Psi^P_1(p^0_1)P_\kappa(p_1)P_\kappa(p_2) \hat{V}_{12;56},\label{planar1}\\
\Gamma^P_2(p_3,p_4,p_5,p_6):=&\ \frac{1}{(2\pi)^8}\hat{V}_{34;56}\times\int \frac{d^4p_1}{(2\pi)^4}\frac{d^4p_2}{(2\pi)^4}\ \Psi^P_2(p^0_1)P_\kappa(p_1)P_\kappa(p_2) \hat{V}_{14;52},\label{planar2}
\end{align}
where the functions $\Psi^P_j$, $j=1,2$, depend only on the external momenta and are given by
\begin{eqnarray}
\Psi^P_1(p^0_1)&=&a_1+b_1e^{3p^0_1/\kappa}+
g_2^2e^{6p^0_1/\kappa}\label{psi1},\\
\Psi^P_2(p^0_1)&=&a_2+b_2e^{3p^0_1/\kappa}+g_1^2e^{3(p^0_3-p^0_1)/\kappa}\label{psi2},
\end{eqnarray}
with
\begin{eqnarray}
a_1&=&2g_1^2(1+e^{3(p^0_3-p^0_4)/\kappa})+g_1g_2(1+e^{3(p^0_3-p^0_5)/\kappa}+2e^{3(p^0_3-p^0_6)/\kappa})e^{3p^0_6/\kappa}\nonumber\\
&+&g_2^2e^{3(p^0_3+p^0_6)/\kappa}\label{a1},\\
b_1&=&g_1g_2(3+e^{3(p^0_3-p^0_4)/\kappa})+2g_2^2e^{3p^0_3/\kappa}\label{b1}\\
a_2&=&2g_1(g_1+g_2e^{3p^0_3/\kappa})+g_1g_2(e^{3p^0_3/\kappa}+e^{3p^0_6/\kappa})\label{a2}\\
b_2&=&2g_2(g_1+g_2e^{3p^0_3/\kappa})+g_2^2e^{3p^0_6/\kappa}+(g_1+g_2e^{3p^0_3/\kappa})(g_1+g_2e^{3p^0_5/\kappa})e^{-3p^0_4/\kappa}\label{b2}
\end{eqnarray}

The non-planar contributions can be written as
\begin{align}
\Gamma^{NP}_1(p_3,p_4,p_5,p_6):=&\ \frac{1}{(2\pi)^8}\int \frac{d^4p_1}{(2\pi)^4}\frac{d^4p_2}{(2\pi)^4}\ \Psi^{NP}_1(p^0_1)P_\kappa(p_1)P_\kappa(p_2)\hat{V}_{14;32}\hat{V}_{56;21},\label{nonplanar1}\\
\Gamma^{NP}_2(p_3,p_4,p_5,p_6):=&\ \frac{1}{(2\pi)^8}\int \frac{d^4p_1}{(2\pi)^4}\frac{d^4p_2}{(2\pi)^4}\ \Psi^{NP}_2(p^0_1)P_\kappa(p_1)P_\kappa(p_2)\hat{V}_{16;24}\hat{V}_{52;31},\label{nonplanar2}
\end{align}
with
\begin{eqnarray}
\Psi^{NP}_1(p^0_1)&=&a_3+b_3e^{3p^0_1/\kappa}+c_3e^{6p^0_1/\kappa}\label{psiNP1},\\
\Psi^{NP}_2(p^0_1)&=&a_4+b_4e^{3p^0_1/\kappa}+c_4e^{6p^0_1/\kappa},\label{psiNP2}
\end{eqnarray}
where
\begin{eqnarray}
a_3&=&g_1^2(1+e^{3(p^0_5-p^0_6)/\kappa})+g_1g_2e^{3p^0_5/\kappa}\label{a3},\\
b_3&=&g_1^2(e^{-3p^0_3/\kappa}+e^{-3p^0_4/\kappa})+g_2^2(e^{3p^0_5/\kappa}
+e^{3p^0_6/\kappa})+g_1g_2(2+e^{3(p^0_5-p^0_4)/\kappa}\nonumber\\
&+&e^{3(p^0_3-p^0_6)/\kappa}+e^{3(p^0_5-p^0_6)/\kappa})\label{b3},\\
c_3&=&g_1g_2(1+e^{-3p^0_4/\kappa})+g_2^2e^{3(p^0_5-p^0_6)/\kappa}\label{c3}
\end{eqnarray}
and
\begin{eqnarray}
a_4&=&g_1(g_1+g_2e^{p^0_5/\kappa})\label{a4},\\
b_4&=&g_1^2e^{-3p^0_3/\kappa}+2g_1g_2+g^2_2e^{3p^0_5/\kappa}\label{b4},\\
c_4&=&g_2(g_2+g_1e^{-3p^0_3/\kappa})\label{c4}.
\end{eqnarray}
\\
We will now show that these contributions are all UV finite.

\subsection{Planar contributions}

Consider first the planar contributions \eqref{planar1}, \eqref{planar2}. It is convenient to use the momentum conservation law stemming from  the vertex functions appearing in each integral of  \eqref{planar1}, \eqref{planar2} to express the internal momentum $p_2$ in term of the other momenta. Namely, the integration over $p_2$ gives rise to
\begin{equation}
p^0_2=p^0_1+Q_j^0, \ \ \vec{p}_2= A_j\left(\vec{p}_1+y^{B_j}\vec{Q}_j\right),\label{variab-interm}
\end{equation}
in which $j=1,2$ refers respectively to $\Gamma^P_1$ and $\Gamma^P_2$. \\
In \eqref{variab-interm}, $Q_j^0$ and $\vec{Q}_j$ are functions of the external momenta (while independent of $y$) which can be easily read off from the vertex functions involving internal momenta in \eqref{planar1}, \eqref{planar2} (recall \eqref{vertex}), while
\begin{eqnarray}
(A_1,B_1)&=&(1,1)\nonumber\\
(A_2,B_2)&=&(e^{-Q_2^0/\kappa},0),
\end{eqnarray}
and in \eqref{variab-interm} we set
\begin{equation}
y=e^{-p^0_1/\kappa}\label{variab-interm2}.
\end{equation}
\\
The UV finiteness of $\Gamma^P_1$ and $\Gamma^P_2$ can be shown in a way somewhat similar to the analysis carried out in \cite{JCW2016}, by exploiting a particular estimate for the propagator $P_\kappa(p)$. Indeed, the following estimate
\begin{equation}\label{borne}
P^{eq}_{\kappa}(q)\leq\frac{e^{-2q^0/\kappa}}{1+e^{-3q^0/\kappa}} \  \frac{8\kappa^2}{\left(\vec{q}^{\hspace{2pt}2}+\kappa^2\mu^2_{-}(q^0)\right)^2},
\end{equation}
holds true, which permits one to control the growth of each of the 2 contributions.\\

We now set:
\begin{equation}
\Gamma^P_j(p_3,p_4,p_5,p_6)=\frac{1}{(2\pi)^8}\hat{V}_{34;56}\times \mathcal{I}^P_j(g_1,g_2;p_3,p_4,p_5,p_6), \ \j=1,2\label{integral-4point}
\end{equation}
where $\mathcal{I}_j$, $j=1,2$ can be read off from \eqref{planar1}, \eqref{planar2}.\\
The analysis of the planar contributions can then be easily carried out by making use of the bound \eqref{borne} together with \eqref{formule1} and \eqref{formule2}. This leads to the following estimates for the amplitudes \eqref{planar1} and \eqref{planar2}, 
\begin{align}
\mathcal{I}^P_j(g_1,g_2;p_3,p_4,p_5,p_6)&\leq\ \frac{e^{-2Q^0_j/\kappa}}{A_j(2\pi)^{10}}\int\frac{y^6\Psi^P_j(y)\ dy}{(1+y^3)(1+y^3e^{-3Q^0_j/\kappa})}\nonumber\\
&\times\int_0^1\frac{x(1-x)\ dx}{\left(-\alpha_j(y)x^2+\beta_j(y)x+\mu_{-}^2(y)\right)^{5/2}}\label{bound-interm},
\end{align}
for $j=1,2$, with
\begin{eqnarray}
\alpha_j(y)&=&\frac{\vec{Q}_j^{\hspace{2pt}2}}{\kappa^2}\ y^{2B_j},\label{fonctionalpha}\\ 
\beta_j(y)&=&\alpha_j(y)+A_j^{-2}\mu_-^2(ye^{-Q^0_j/\kappa})-\mu_-^2(y),\label{fonction1}
\end{eqnarray}
and $\Psi^P_j(y)$ still given by \eqref{psi1}, \eqref{psi2}.\\

The second integral in the RHS of \eqref{bound-interm} can be easily bounded to give rise to
\begin{align}
\mathcal{I}^P_j(g_1,g_2;p_3,p_4,p_5,p_6)&\leq\ \frac{e^{-2Q^0_j/\kappa}}{6A_j(2\pi)^{10}}\int\frac{y^6\Psi^P_j(y)\ dy}{(1+y^3)(1+y^3e^{-3Q^0_j/\kappa})}\nonumber\\
&\times\max\left[\frac{1}{\mu^2_{-}(y)},\frac{A^2_j}{\mu^2_{-}(ye^{-Q^0_j/\kappa})}\right]^{5/2},\label{eq-planar}
\end{align}
for $j=1,2$.\\

By inspection, one easily finds that the integrand of the RHS of \eqref{eq-planar} for $\Gamma^P_1$ ($j=1$) behaves as $\sim y^3$ for $y\to0$ (resp. $\sim y^{-5}$ for $y\to\infty$ while for $\Gamma^P_2$ ($j=2$) it behaves as a constant for $y\to0$ (resp. $\sim y^{-2}$ for $y\to\infty$). \\
Hence, the RHS of \eqref{eq-planar} is always finite for $j=1,2$. From this, one concludes that $\Gamma^P_1$ and $\Gamma^P_2$ are UV finite.\\

\subsection{Non-planar contributions}

Consider now the non-planar contributions $\Gamma_j^{NP}$, \eqref{nonplanar1}, \eqref{nonplanar2}. We start with \eqref{nonplanar1}. To deal with the integrals over internal momenta, it is convenient to integrate first over $p_2$. This yields
\begin{equation}
p^0_2=p^0_1+(p^0_6-p^0_5),\ \ \vec{p}_2=\vec{c}_1y+\vec{c}_2,\ \ \vec{p}_1=\vec{c}_3y+\vec{c}_4,\label{NP1p2},
\end{equation}
in which the quantities $c_j$, $j=1,...,4$ depend only on the external momenta. Their respective expressions which can be easily obtained from a mere computation using the vertex functions will not be needed in the ensuing analysis.\\

Then, by further using \eqref{NP1p2} in \eqref{nonplanar1} together with the bound \eqref{borne}, one finds that the non-planar contributions can be conveniently cast into the form
\begin{equation}
\Gamma^{NP}_j(p_3,p_4,p_5,p_6)=\delta\left(p_4^0-p_3^0+p_6^0-p_5^0\right)\times\mathcal{I}^{NP}_j(g_1,g_2;p_3,p_4,p_5,p_6), \ \ j=1,2\label{integral-4point-prime}
\end{equation}
in which $\mathcal{I}^{NP}_j(g_1,g_2;p_3,p_4,p_5,p_6)$ is bounded.\\
For $\Gamma^{NP}_j$ ($j=1$), one finds the following estimate
\begin{align}\label{eqNP1}
\mathcal{I}^{NP}_1(g_1,g_2;p_3,p_4,p_5,p_6)&\le\text{C}_1\int dy\ \frac{\kappa^5
y^9\Psi_1^{NP}(y)\ }{(1+y^3)(1+c_0^3y^3)}\nonumber\\
&\times\frac{1}{\left[(\vec{c}_3y+\vec{c}_4)^2+\kappa^2\mu_-^2(y)\right]^2\left[(\vec{c}_1y+\vec{c}_2)^2+\kappa^2\mu_-^2(c_0y)\right]^2},
\end{align}
where $C_1$ is a constant collecting all the unessential factors and $c_0$ depends only on the external momenta.\\
In the same way, the integration over $p_2$ in \eqref{nonplanar2} yields
\begin{equation}\label{NP2p2}
p^0_2=-p^0_1+(p^0_3+p^0_5),\ \ \vec{p}_2=\frac{1}{y}\left(\vec{c}_1^{\hspace{3pt}'}y+\vec{c}_2^{\hspace{3pt}'}\right),\ \ \vec{p}_1=\vec{c}_3^{\hspace{3pt}'}y+\vec{c}_4^{\hspace{3pt}'},
\end{equation}
where the quantities $c^\prime_j$, $j=1,...,$4 again depend only on the external momenta. From \eqref{NP2p2} and the bound \eqref{borne}, one infers
\begin{align}\label{eqNP2}
\mathcal{I}^{NP}_2(g_1,g_2;p_3,p_4,p_5,p_6)
&\le\text{C}_2\int dy\ \frac{\kappa^5y^9\Psi_2^{NP}(y)\ }{(1+y^3)(y^3+c_0^{'\hspace{2pt}3})}\nonumber\\
&\times\frac{1}{[(\vec{c}_4^{\hspace{3pt}'}+\vec{c}_3^{\hspace{3pt}'}y)^2+\kappa^2\mu_-^2(y)]^2[(\vec{c}_2^{\hspace{3pt}'}+\vec{c}_1^{\hspace{3pt}'}y)^2+
\kappa^2y^2\mu_-^2(\frac{c_0^{'}}{y})]^2},
\end{align}
where again $C_2$ is a constant and $c^\prime_0$ depends only on the external momenta. By inspection of the integrals in \eqref{eqNP1} and \eqref{eqNP2}, one realizes that both expressions are finite, thus checking that the non-planar contributions are UV finite for generic non exceptional external momenta as it could have been expected.\\

Hence, we find that the 1-loop corrections to the 4-point function in the present orientable scalar NCFT on $\kappa$-Minkowski space are UV finite.\\

\section{Discussion and conclusion}\label{section5}

We have considered a family of $\kappa$-Poincar\'e invariant scalar NCFT on 4-d $\kappa$-Minkowski space $S_\kappa(\phi,\phi^\dag)$ with orientable interaction, i.e for which $\phi$ and $\phi^\dag$ alternate in the quartic interaction term. The kinetic operator was assumed to be the square of the $U_\kappa(iso(4))$-equivariant Dirac operator built in \cite{dandrea2006}. This family depends on 2 real dimensionless coupling constants $g_1$ and $g_2$ and a parameter with dimension of a mass and reduces in the formal commutative limit $\kappa\to\infty$ to the standard complex $\phi^4$ theory. The action functional involves a star product related to a Weyl-type quantization which describes the $\kappa$-deformation of the Minkowski space together with a twisted trace implied by the $\kappa$-Poincar\'e invariance. These can be conveniently re-expressed as to represent the action functional as a non local (albeit commutative) action functional depending on the twist operator defining the trace, which appears to drastically controls the UV behavior of the quantum fluctuations. The relationship of the twist with the Tomita modular operator for the KMS condition which holds true on the algebra of the fields of the present NCFT has been discussed in \cite{PW2018}.\\

Thanks to the decay properties of the propagator at large momenta, the actual computation of the 1-loop contributions to the 2- and 4- point functions reduces to the control of time-like integrals. These latter can be dealt with by using a convenient regulator inspired by the algebraic expressions for the generators of the deformed translation algebra, a subalgebra of the Hopf $\kappa$-Poincar\'e algebra.\\
We have found that the 2-point function receives UV linearly diverging 1-loop corrections, namely 
\begin{equation}
\omega_j\sim g_j\kappa\Lambda_0+\text{finite}, \label{mass-div}
\end{equation}
for $j=1,2$, where $\Lambda_0$ is the cut-off, which is milder than its counterpart for the $\phi^4$ commutative limit. Besides, we have found that the 2-point function stays free of IR singularities at exceptional momenta whose occurrence would have signaled the presence of UV/IR mixing in the NCFT. Note that in view of \eqref{mass-div}, the form of the mass term $\int d^4p\ \bar{\phi}(p)\phi(p)m^2(1+e^{-3p_0/\kappa})$ is preserved by the radiative corrections only when $g_1=g_2$, which corresponds to the invariance of the classical action functional \eqref{kaction}, \eqref{kinetic-term}, \eqref{interaction} under $\phi\leftrightarrow\phi^\dag$.\\
We have found that within the present orientable NCFT, all the 1-loop planar and non-planar contributions to the 4-point function are UV finite. This results from the existence of the particular estimate \eqref{borne} satisfied by the propagator combined in part with the decay properties of the propagator at large momenta.\\

A somewhat similar analysis can be performed by replacing the kinetic operator \eqref{equi-dirac} by the Casimir operator for the $\kappa$-Poincar\'e algebra \eqref{Casimir}. One would then find higher degrees of UV divergence for the 2-point and 4-point functions (increased by up to 2 unit compared to the present case) leading to an overall behavior for the corresponding NCFT which is qualitatively similar to the one of the commutative $\phi^4$ commutative model. \\

As it has been shown above, the situation for the present orientable NCFT is different since only the mass terms have to be renormalized while the coupling constants receive only finite corrections and there is no wave function renormalization. Since the coupling constant receive only finite renormalisation, they will not depend on a (mass) scale at the 1-loop order. In other words, the coupling constants are scale-invariant at the 1-loop order which signals that the beta functions for the coupling constants are zero at this order.

\vskip 2 true cm

{\bf{Acknowledgments:}} One of us (J.-C.W) thanks F. D'Andrea for correspondence on spectral triples related to the present framework. We also thank L. Freidel, T. Juri\'c and D. Vassilevich for useful discussions and comments.

{\small%

}%
\end{document}